\documentclass[letter]{aa}
\usepackage[fleqn]{amsmath}
\usepackage{epstopdf}
\usepackage{multirow}
\usepackage{graphicx}
\usepackage{txfonts}
\usepackage{expdlist}
\usepackage{natbib}
\usepackage{color}
\usepackage{epsfig}
\usepackage{pdflscape}
\usepackage{lscape}
\usepackage{rotating}

\newcommand{\grau}   	{$^{\circ}$} 
\newcommand{\Mm}		{$\pm$}

\newcommand{\LT}		{$\le$}

\newcommand{\mJy}		{~mJy~beam$^{-1}$}
\newcommand{\Jy}		{~Jy~beam$^{-1}$}
\newcommand{\msun}	    {~M$_{\sun}$\xspace}

\newcommand{\cmd}		{~cm$^{-2}$\xspace}

\newcommand{\ie}        {i.\,e.,}
\newcommand{\eg}    	{e.\,g.,}

\newcommand{\phn}		{\phantom{0}}
\newcommand{\phnn}		{\phantom{00}}

\begin{document}

\title{Unveiling a cluster of protostellar disks around the massive protostar GGD\,27 MM1}
\titlerunning{Protostellar disks around the massive protostar GGD\,27 MM1}

\author{ 
G.~Busquet\inst{1,2} 
\and
 J.~M.~Girart\inst{1,2} 
\and
R.~Estalella\inst{3} 
\and
M.~Fern\'andez-L\'opez\inst{4} 
\and
R.~Galván-Madrid\inst{5}
\and
G.~Anglada\inst{6}
\and
C.~Carrasco-Gonz\'alez\inst{5} 
\and
N.~Añez-López\inst{1}
\and
S.~Curiel\inst{7} 
\and
M.~Osorio\inst{6}
\and
L.~F.~Rodríguez\inst{5}
\and
J.~M.~Torrelles\inst{1,2}
}
\authorrunning{Busquet et al.}

\institute{
Institut de Ci\`encies de l'Espai (ICE, CSIC), Can Magrans, s/n, E-08193 Cerdanyola del Vall\`es, Catalonia \\
\email:{[busquet,girart]@ice.cat}
\and
Institut d'Estudis Espacials de Catalunya (IEEC), E-08034, Barcelona, Catalonia 
\and
Departament de F{\'\i}sica Qu\`antica i Astrof{\'\i}sica, Institut de Ci\`encies del Cosmos (ICC), Universitat de Barcelona (IEEC-UB), Mart{\'\i} i Franqu\`es 1, E-08028 Barcelona, Catalonia, Spain
\and
Instituto Argentino de Radioastronom{\'\i}a, CCT-La Plata (CONICET), C.C.5, 1894, Villa Elisa, Argentina
\and
Instituto de Radioastronom\'ia y Astrof\'isica (IRyA), UNAM, Apdo. Postal 72-3 (Xangari), Morelia, Michoac\'an 58089, Mexico
\and
Instituto de Astrofísica de Andalucía (IAA, CSIC), Glorieta de la Astronomía, s/n, E-18008 Granada, Spain
\and
Instituto de Astronom{\'\i}a, Universidad Nacional Aut\'onoma de M\'exico (UNAM), Apartado Postal 70-264, 04510 M\'exico, DF, M\'exico
}

\date{ Received {\it date} / Accepted {\it date} }

\abstract{
Most stars form in clusters, and thus it is important to characterize the protostellar disk population in dense environments 
to assess whether the environment plays a role in the subsequent evolution; specifically, whether planet formation is altered with respect to more isolated stars formed in dark clouds. 
} 
{
Investigate the properties of the protostellar disks in the GGD\,27 cluster 
and compare them with those obtained from disks formed in nearby regions. } 
{
We used ALMA to observe the star-forming region GGD\,27 at 1.14~mm with an unprecedented angular resolution, 40~mas ($\sim56$~au)  
and sensitivity ($\sim0.002$\msun).}
{

We detected a cluster of 25 continuum sources, most of which are likely tracing disks around Class~0/I protostars. Excluding the two most massive objects, disks masses are in the range 0.003--0.05\msun. The analysis of the cluster properties indicates that GGD\,27 displays moderate subclustering. This result combined with the dynamical timescale of the radio jet ($\sim10^4$~years) suggests the youthfulness of the cluster. The lack of disk mass segregation signatures may support this too.
We found a clear paucity of disks with $R_{\rm{disk}}>100$~au.
The median value of the radius is 34~au, smaller than the median of 92~au for Taurus but comparable to the value found in Ophiuchus and in the Orion Nebula Cluster. In GGD\,27 there is no evidence of a distance-dependent disk mass distribution (\ie\ disk mass depletion due to external photoevaporation), most likely due to the cluster youth.

There is a clear deficit of disks for distances $<0.02$~pc. Only for distances $>0.04$~pc stars can form larger and more massive disks, suggesting that dynamical interactions far from the cluster center are weaker, although the small disks found could be the result of disk truncation. This work demonstrates the potential to characterize disks from low-mass YSOs in distant and massive (still deeply embedded) clustered environments. }
{}


\maketitle

\section{Introduction}
One of the major challenges of modern astrophysics is the direct observation of protostellar disks, the progenitors of planetary systems, in all stages of star formation, from extremely young protostars, still deeply embedded in the natal dense envelope, to more evolved pre-main sequence stars, at which point the envelope has dissipated and a young star (or stars) and a protoplanetary disk remain. 
The advent of new facilities such as the Atacama Large Millimeter/submillimeter Array (ALMA) and the Jansky Very Large Array (JVLA) have boosted the study of the formation and evolution of disks around young solar-type stars and started to unveil the details of the planetary formation process \citep[\eg][]{Andrews2016,Carrasco2016}.
Extensive surveys conducted mainly toward nearby and isolated low-mass star-forming regions, generally unaffected by their external environment, 
lead to a major progress in the understanding of the disk properties and its evolution \citep[\eg][]{Andrews2016b}. Although some fraction of the stellar population form in an isolated mode such as in the Taurus-Auriga star-forming region, most stars, including our Sun, formed in clusters that are embedded within molecular clouds \citep[\eg][]{Lada2003,Adams2010}. In cluster environments, the high stellar density, protostellar outflow feedback, and ultraviolet (UV) radiation from the massive star(s) can influence the disk properties and evolution of the low-mass cluster members like in the Orion Nebula Cluster (ONC), where the extreme UV radiation field inhibits the potential to form planets close to the massive stars \citep{Mann2014}. To date, only a few publications reporting disks properties in clustered environments can be found in the literature for Orion \citep[\eg][]{Mann2015,Eisner2016,Eisner2018},
the Upper Scorpius OB Association \citep[\eg][]{Carpenter2014,Barenfeld2017}, and the IC\,348 cluster \citep[\eg][]{Ruiz-Rodriguez2018}. However, most of these studies focus on relatively evolved clusters.
Characterizing the disk population in the earliest stages of star formation in cluster environments is essential to understand the initial conditions, the formation process, and the final properties of young planets.

GGD\,27 is a reflection nebula at a distance of 1.4~kpc (Girart et al. in prep.)\footnote{Based on GAIA-DR2 astrometry of a sample of pre-main sequence stars identified using near-infrared imaging polarimetry data of GGD\,27 \citep{Kwon16}.}.  
It is illuminated mainly by a massive B0-type young stellar object (YSO), IRAS\,18162$-$2048, that powers a long (14~pc), fast and  highly collimated radio jet associated with HH 80-81 \citep[\eg][]{Marti93,Masque12}.
The high-mass YSO driving the HH80-81 jet is associated with the millimeter source MM1, a massive dusty disk surrounded by a warm and dense elongated molecular envelope \citep[\eg][]{Girart17, Girart2018}.  
There is another site of massive star formation, the core MM2, located $7''$ ($\sim10000$~au) northeast of MM1. 
These two massive protostellar sites are surrounded by a cluster of infrared and X-ray sources \citep[\eg][]{Aspin94, Pravdo04, Pravdo09, Qiu08}.

In this letter we report the detection of a cluster of 1.14~mm continuum sources, most of them previously unknown. The data is part of the polarimetric ALMA observations towards GGD\,27 MM1 presented in \citet{Girart2018}. A brief description of the observations is given in Sect.~\ref{Sec:Obs}. The results are presented in Sect.~\ref{Sec:Res}. We analyze and discuss the properties of the cluster in comparison with previous works in other regions in Sect.~\ref{Sec:Discus}. 

\section{Observations}\label{Sec:Obs}

The observations were performed on 2015 December in the C36-8/7 configuration at the central frequency of 263~GHz (1.14~mm). A detailed description is given in \citet{Girart2018}. In order to highlight the emission of compact and weak sources within the field of view of $22\farcs1$ at 263~GHz, we performed maps with a robust weighting of 0.5 and excluded baselines shorter than 300~k$\lambda$. The robust weighting allows to obtain a good compromise between sensitivity and angular resolution. The angular resolution achieved is 45.0~mas$\times$38.3~mas with a position angle of $-62.4^{\circ}$. The exclusion of baselines shorter than 300~k$\lambda$ allows to have higher fidelity cleaned images \citep{Girart2018}. The images presented here are corrected by the primary beam attenuation. 
The sensitivity is not uniform along the field of view. First, the correction of the primary beam attenuation makes the rms noise to increase outwards with respect to the phase center. Additionally, the presence of extended emission from the  envelope around MM2, and specially around MM1, 
increases the rms noise nearby these sources. Thus, we adopted an 8$\sigma$ threshold for source detection, where $\sigma$ is the local rms noise measured within a radius of $0\farcs5$ of the source and excluding its emission. Table~\ref{tsources} shows the source name, position, peak intensity, flux density, brightness temperature, and counterpart identification. We include three additional tentative sources detected at 5$\sigma$ that are located in regions without signatures of extended emission.

\begin{figure}[!t]
\begin{center}
\epsfig{file=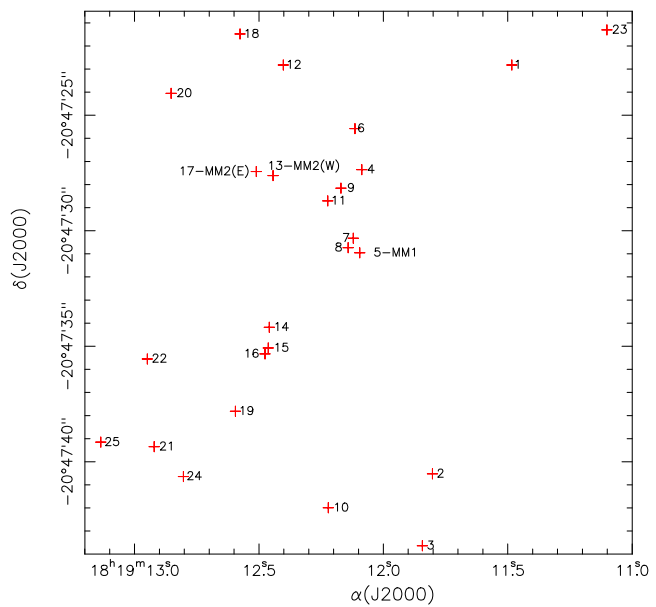,scale=0.45,angle=0}
\caption{Positions (indicated by crosses and labeled according to Table~\ref{tsources}) of the dust continuum sources detected in the GGD\,27 region.  
Images of the individual sources are shown in Fig.~\ref{fig:Sources}.
}
\label{fig:FOV}
\end{center}
\end{figure}

\section{Results}\label{Sec:Res}

\begin{figure}[htpb]
\begin{center}
\epsfig{file=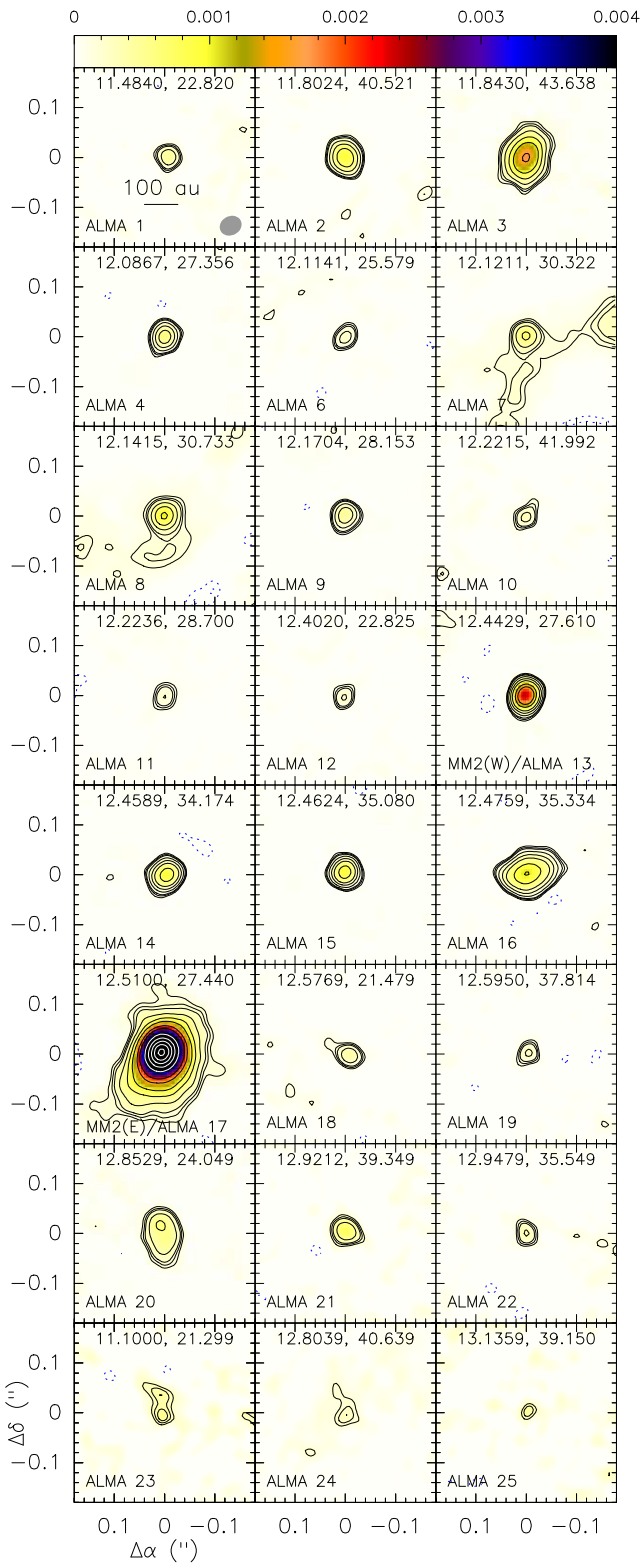,scale=0.71,angle=0}
\caption{Composite of the all sources detected in the field of view (except for MM1/ALMA\,5). The images have been corrected for the primary beam response. Black contours are $-4$, $-3$, 3, 4, 5, 8, 13, 20, 35 and 50 times the local rms noise of each image (Table~\ref{tsources}). White contours are 100, 150, 200, 250 and 300 times the local rms noise. Table~\ref{tsources} lists their absolute positions. In each panel we labeled the second and arcsecond position of the center of the panel.
The color scale is the same for all panels and it is shown at the top of the figure (units of \Jy). The synthesized beam is shown in the bottom-right corner of the first image.}
\label{fig:Sources}
\end{center}
\end{figure}

Figure~\ref{fig:FOV} shows the position of the 1.14~mm continuum sources detected with ALMA.  We detected a cluster of 25 continuum sources, including the three previously know sites of massive star formation:  MM1/ALMA\,5,  MM2(W)/ALMA\,13 and MM2(W)/ALMA\,17 \citep[][here on MM1, MM2(W) and MM2(E), respectively]{FernandezC11, Girart17}. The remaining 22 sources are new detections (Table~\ref{tsources}) and are distributed mainly north and southeast of MM1. In Fig.~\ref{fig:Sources} we present close-up images of all ALMA detections, including the three tentatively detected sources. Most of the continuum sources are compact, and only two, ALMA\,7 and ALMA\,8, which are located close to MM1, present faint and extended emission.

We assumed that the continuum emission arises from disks surrounding YSOs. This assumption is based on the derived sizes of a few 100~au or smaller (see below) and in the fact that no more than one millimeter extragalactic background is expected in the field of view  (see Appendix~\ref{Appendix:Backrground} for a justification). 

We obtained physical properties (masses, radius, inclination angle and position angle) for the detected disks. Appendix~\ref{Appendix:Radii} describes the method used to derive these properties, which assumes that the intensity can be described with a power-law of the radius. Table~\ref{tdiskpar} shows the disk parameters inferred with this method. 
Except for the two massive disks, MM1 and MM2(E), the millimeter sources have disks masses in the range 0.003--0.05\footnote{We note that the derived masses are most likely a lower limit since we assumed optically thin emission (see Appendix~\ref{Appendix:Radii}).}\msun\ (Table~\ref{tdiskpar}), suggesting that these disks are associated with low-mass YSOs. For these low-mass YSOs, the median value of the disk radius is 34~au and only two YSOs have large disks ($R_{\rm{disk}}\ga 100$~au). 
From Fig.~\ref{fig:FOV}  (see also Fig.~\ref{fmst}b)  we can distinguish three possible multiple systems: the already known binary system formed by MM2(E) and MM2(W) whose separation is $\sim$1400~au, and two triple systems formed by MM1/ALMA\,7/ALMA\,8 and ALMA\,14/ALMA\,15/ALMA\,16, separated by $\sim$1000~au and $\sim$1600~au, respectively.

\begin{table}
\caption{Disk Properties in the GGD\,27 protostellar cluster.}
\begin{tabular}{@{\hspace{-1mm}}l@{\hspace{-8mm}}r@{\hspace{1mm}}r@{\hspace{1mm}}r@{\hspace{1mm}}r@{\hspace{1mm}}r}

\hline
&
\multicolumn{1}{c}{$M_{\rm disk}^a$} &
\multicolumn{2}{c}{$R_{\rm disk}$} &
\multicolumn{1}{c}{$i^b$} &
\multicolumn{1}{c}{$PA^b$} 
\\
\cline{3-4}
Name & 
\multicolumn{1}{c}{($10^{-3} \, \rm{M}_{\odot}$)} &
\multicolumn{1}{c}{(mas)} &
\multicolumn{1}{c}{(au)} &
\multicolumn{1}{c}{(\grau)} &
\multicolumn{1}{c}{(\grau)} 
\\
\hline 
ALMA 1  			& 6		
& \LT17 	& \LT24		& --  		& -- 			\\
ALMA 2  			& 18
&38\Mm6 	& 53\Mm8	& 60\Mm15	& $-$45\Mm10	\\
ALMA 3  			& 54	
& 68\Mm3 	& 95\Mm4
& 41\Mm4 	& 55\Mm3 		\\
ALMA 4  			& 8	
& 20\Mm10 	& 28\Mm14	& 67\Mm19	& 31\Mm31 		\\
MM1/ALMA 5			&574  
& 171\Mm6	& 239\Mm8	& 49\Mm1	& 113\Mm1  		\\
ALMA 6  			& 3		
& \LT27 	& \LT38 	& --		& --			\\
ALMA 7				& 12	
& 23\Mm13	& 32\Mm18	& 0   		& 80\Mm39		\\
ALMA 8				& 17	
& 20\Mm5	& 28\Mm7	& 0 		& --			\\
ALMA 9  			& 7	
& 14\Mm8	& 20\Mm11	& 0 		& -- 			\\
ALMA 10  			& 4		
& 39\Mm18 	& 55\Mm25 	& 84\Mm15 	& 42\Mm26		\\
ALMA 11  			& 4		
& 38\Mm16	& 53\Mm22	& 0			& --   			\\	
ALMA 12  			& 3		
& 15\Mm3 	& 21\Mm4	
& 76\Mm5 	& 32\Mm12 		\\
MM2(W)/ALMA 13		&16		
& 19\Mm2 	& 27\Mm3 	& 60\Mm30 	& 17\Mm11 		\\
ALMA 14  			& 13	
& 25\Mm4 	& 35\Mm6	& 80\Mm6 	& 18\Mm5		\\
ALMA 15   			& 12	
& 16\Mm5 	& 22\Mm7 	& 72\Mm2 	& 6\Mm3			\\
ALMA 16  			& 32	
& 75\Mm2 	& 105\Mm3 	& 66\Mm2 	& 12\Mm1		\\
MM2(E)/ALMA 17	 	&194	
& 88\Mm1 	& 123\Mm1	& 39\Mm1 	& 40\Mm1	\\
ALMA 18  			& 5	
& \LT23 	& \LT32 	& --		& --			\\
ALMA 19  			& 4		
& \LT22		& \LT31		& -- 		& -- 			\\
ALMA 20  			& 22	
& 73\Mm8 	& 102\Mm11	& 51\Mm8 	& $-67$\Mm10 	\\  
ALMA 21  			& 14	
& 31\Mm9 	& 43\Mm11	& 80\Mm21 	& $-15$\Mm27 	\\ 
ALMA 22  			& 5		
& 25\Mm16 	& 35\Mm22 	& 0 		& -- 			\\
ALMA 23$^c$ 		& 14	
& -- 		& -- 		& -- 		& -- 			\\
ALMA 24$^c$ 		& 5		
& --		& --		& -- 		& -- 			\\
ALMA 25$^c$ 		& 5		
& --		& --		& -- 		& -- 			\\
\hline
\end{tabular}
\\
{\em $^a$} See Appendix~C for more details on how the mass is estimated. 
{\em $^b$} $i$ is the inclination angle, where $i=0$\grau\ for a face-on disk, and $PA$ is the position angle of the disk axis projected on the plane of sky (i.e.\ the ellipse minor axis). \\
{\em $^c$} Source too weak to derive its size and geometry.
\\
\label{tdiskpar}
\end{table}

\section{Analysis and Discussion}\label{Sec:Discus}

\subsection{Clustering and mass segregation} \label{sec:statistics}

\begin{figure}[!t]
\begin{center}
\begin{tabular}[t]{c} 
\epsfig{file=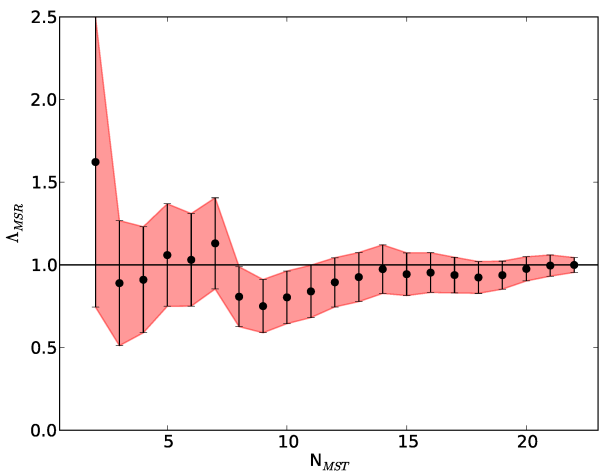,scale=0.45,angle=0} 
	\end{tabular}
\caption{Mass Segregation Ratio ($\Lambda_\mathrm{MSR}$, see Appendix~\ref{appendix:segregation}), as a function of the $N_\mathrm{MST}$ (MST: Minimum Spanning Tree) most massive disks of GGD\,27. ALMA\,1 and ALMA\,2, classified as Class~II and Class~III YSOs, respectively, are not included. The main deviation from a uniform behavior around unity is for $N_\mathrm{MST}=2$, suggesting that the two most massive sources are closer to each other than the average distance between two random sources. The horizontal line at $\Lambda_\mathrm{MSR}=1$ marks the regime where sources show no evidence for mass segregation.}
\label{fcluster}
\end{center}
\end{figure}

In cluster environments, gravitational interactions group the most massive objects toward the center in a relatively short time and generates radial mass segregation and a smooth density distribution of sources that can be approached as a power law 
\citep[\eg][]{2004Cartwright,2009Allison}.
However, the clustering analysis done in Appendix~\ref{appendix:cluster} reveals a fractal type of subclustering with three possible subcenters separated by $\sim0.04$~pc: around MM1, MM2(E)-MM2(W) and the triple system ALMA\,14--16. In addition, the GGD\,27 cluster shows  evidence for a lack of primordial mass segregation signatures in the disk population, but rather a quite random distribution (see Fig.~\ref{fcluster} where we present the mass segregation ratio as a function of the $N_{\mathrm{MST}}$ most massive disks and Appendix~\ref{appendix:segregation}). 
Only the two most massive protostellar disks may have already relaxed toward the center of the cluster, since the relaxation time is inversely proportional to the protostellar system mass. This is in contrast with respect to the Serpens South protocluster that shows evidence of primordial mass segregation \citep{Plunkett2018}.
All these facts combined (\ie\ the low value of the $\bar{Q}$ parameter, the dynamical timescale of the radiojet ($\sim10^4$~years) and the lack of mass segregation signtaures, despites its caveats and uncertainties of this analysis) suggest that the cluster of disks is very young, probably associated with Class~0 and Class~I low-mass protostars (see Appendix~\ref{appendix:segregation} for further discussion). This is in agreement with the properties of the X-ray sources found in the region \citep{Pravdo09}, who reported a cluster of Class~I protostars in GGD\,27.
Interestingly, the two massive YSOs also appear to be in a very early stage of evolution. MM1 powers the HH80-81 radio jet with a dynamical time of $\sim10^4$~years \citep{Masque12}. 
MM2(E) is even in an earlier stage \citep{Qiu09, FernandezC11,Girart17}. This suggests certain coevality of the low-mass cluster members with the two massive YSOs and therefore a smaller age dependece of the protostellar disk masses (see Appendix~\ref{appendix:segregation}). Our analysis indicates that the spatial distribution of the protostellar disk population does not depend on the disk mass.

%--------------------------------------------------------
%-------------------------------------------------------

\subsection{Comparison with nearby regions}
\begin{figure}[!t]
\begin{center}
\begin{tabular}[t]{c} 
\epsfig{file=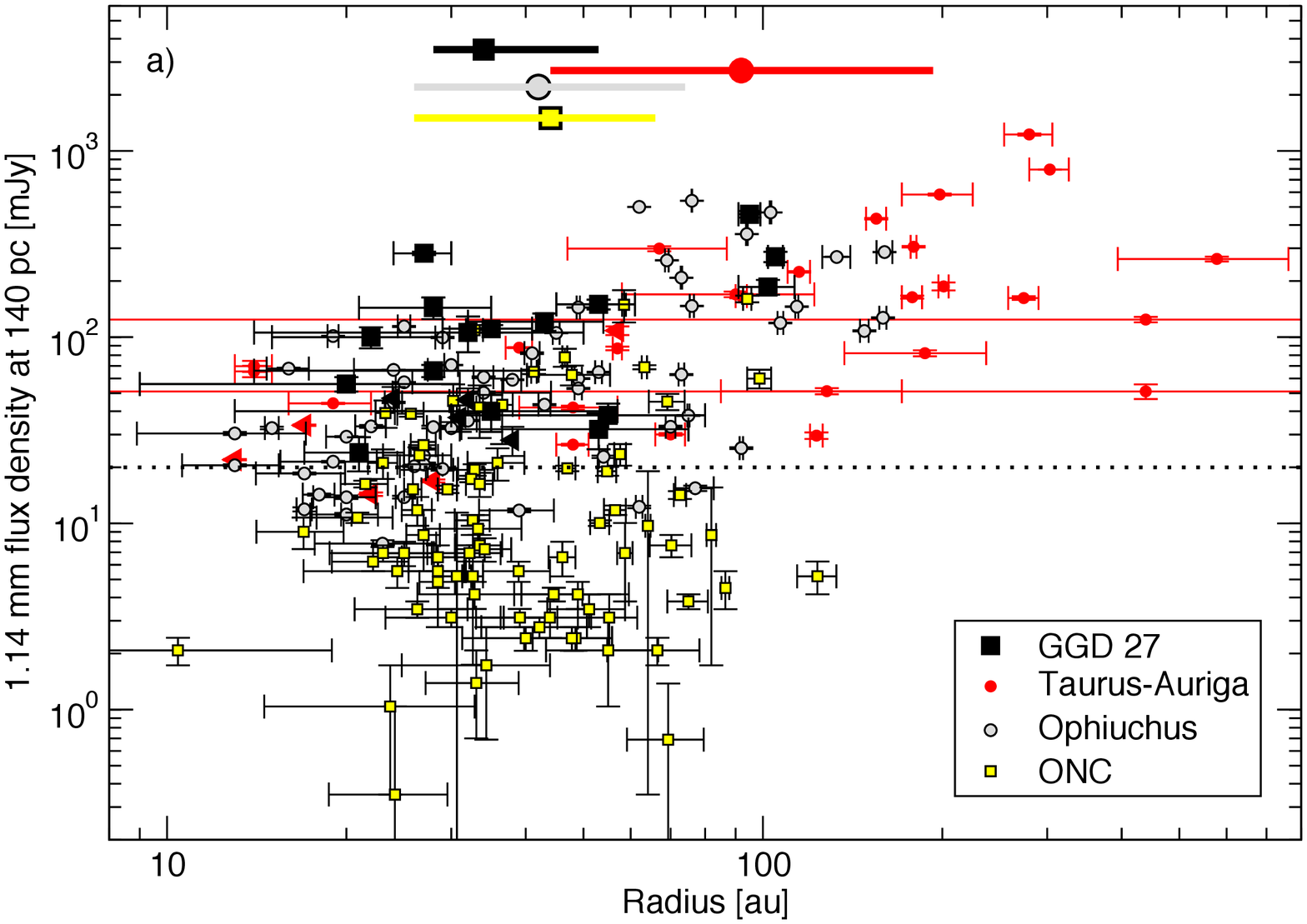,scale=0.32,angle=0} \\
\epsfig{file=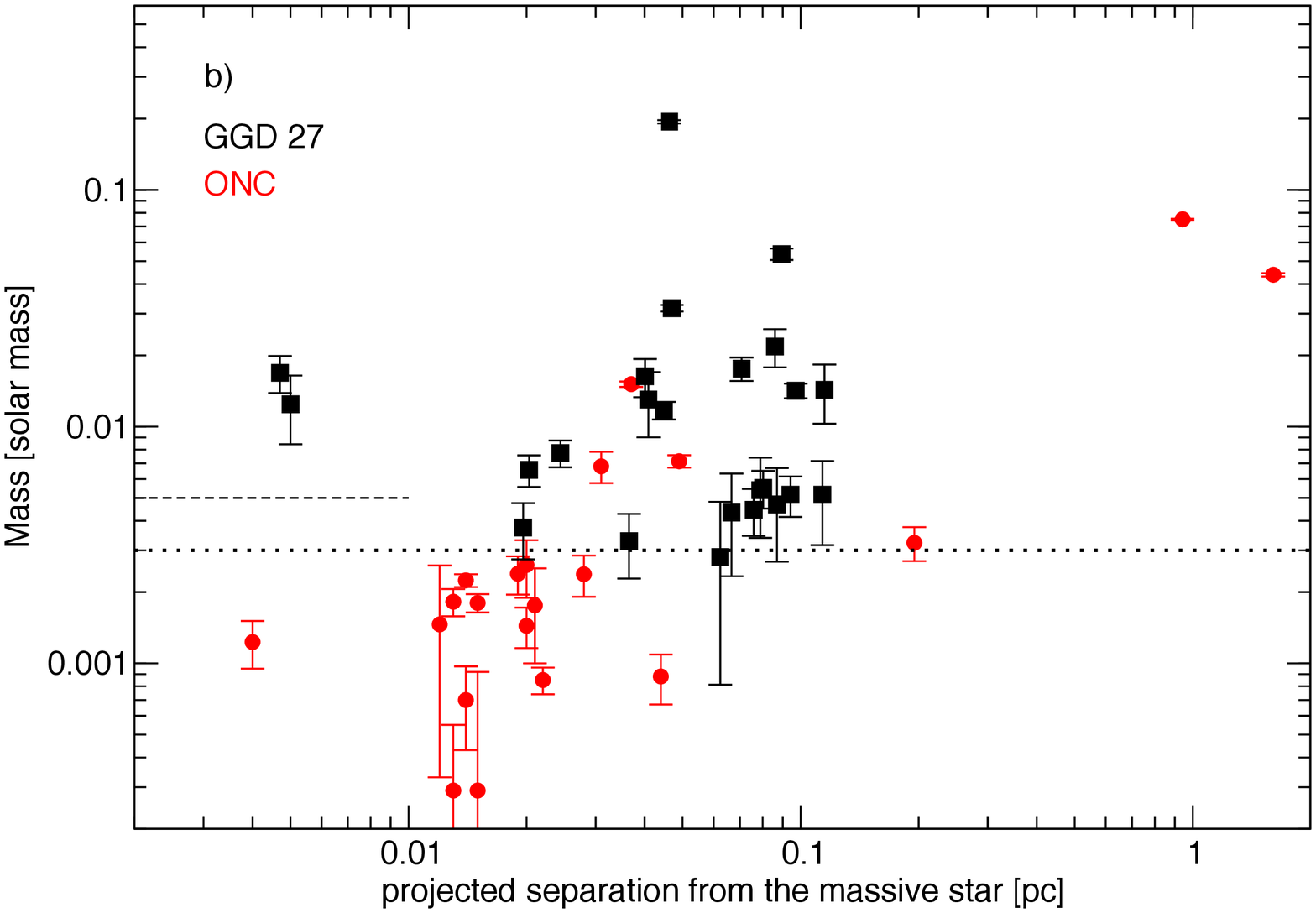,scale=0.32,angle=0}\\
\epsfig{file=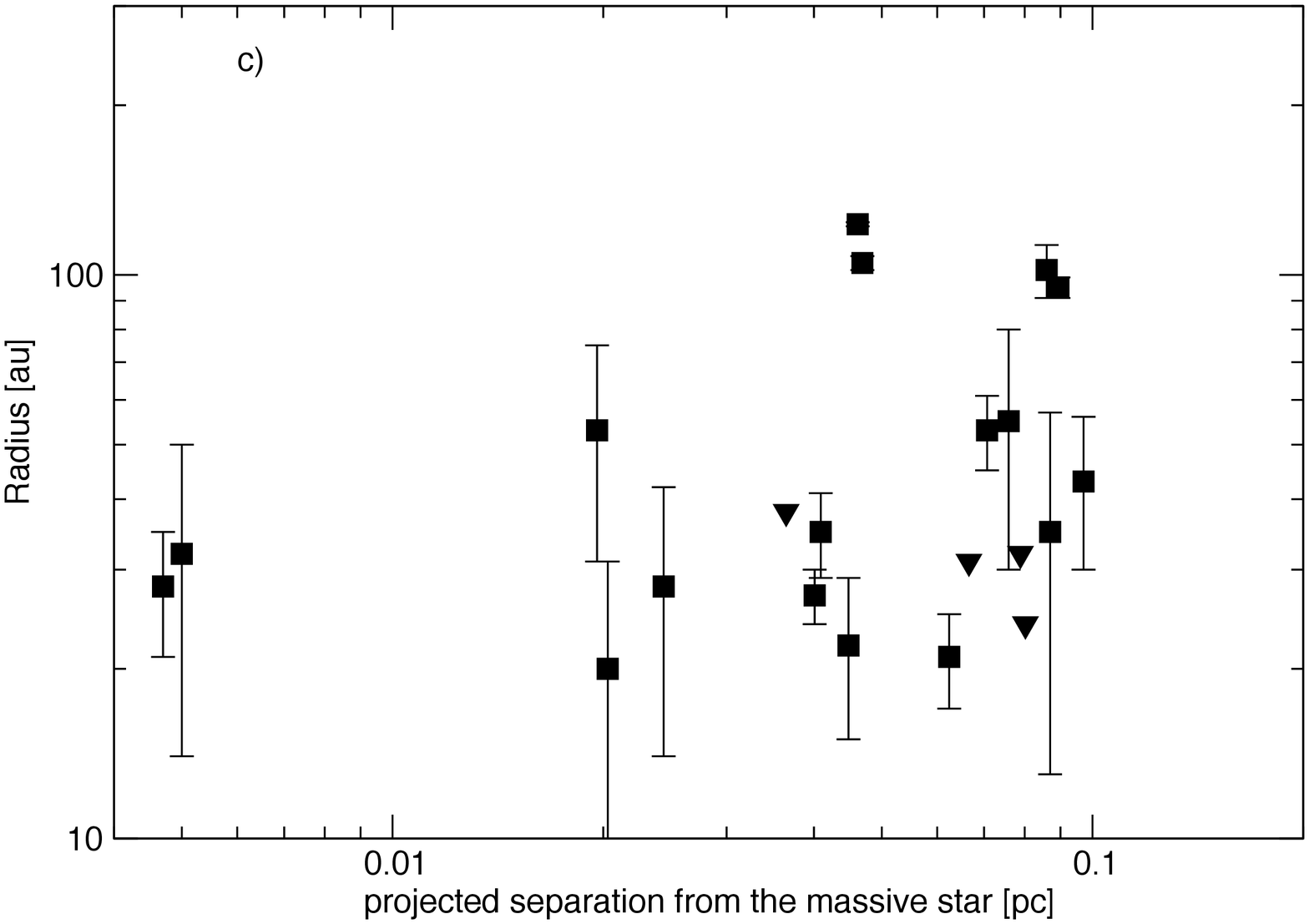,scale=0.32,angle=0}
	\end{tabular}
\caption{\emph{a):} 1.14~mm flux (scaled to a distance of 140~pc) as a function of the radius for all disks detected in GGD\,27 (black squares and triangles), Taurus-Auriga (red circles; \citealt{Guilloteau2011,Pietu2014}), Ophiuchus (grey circles; \citealt{Cieza2019}), and ONC (yellow squares; \citealt{Eisner2018}).  The flux at 1.14~mm was extrapolated from the measurements at 1.3~mm  for Taurus and Ophiuchus disks and from 850~$\mu$m for ONC disks 
using an spectral index of $\alpha=3$.
The dotted line depicts the average mass sensitivity limit ($8\sigma$). We show, in the top, the median and its dispersion (1st and 3rd quartiles) for GGD\,27 (black), Taurus (red), Ophiuchus (grey) and ONC (yellow) samples. 
The most massive disks MM1 and MM2(E) have been removed from the plot.
\emph{b):} Disk masses in GGD\,27 (black squares) and ONC (red circles) clusters as a function of their projected distance from the massive star IRAS\,18162$-$2048 and $\theta^{\rm{1}}$\,Ori\,C \citep{Mann2014}, respectively. The dotted line depicts the average mass sensitivity limit while the dashed line indicates the higher $8\sigma$ limit close to MM1. \emph{c):} Disk radii in GGD\,27 as a function of their projected distance from the massive star. In all panels upper limits in $R_{\rm{disk}}$ are indicated with triangles.
}
\label{fmass}
\end{center}
\end{figure}

In this section, we investigated whether the disk properties of the GGD\,27 cluster display similar/different features than those located in nearby regions. 
Figure~\ref{fmass}a shows the 1.14~mm flux as a function of the radius of the disks for GGD\,27, Taurus-Auriga \citep{Guilloteau2011,Pietu2014}, Ophiuchus \citep{Cieza2019}, and ONC \citep{Eisner2018}. 
We scaled the observed fluxes of GGD\,27 to the distance of 140~pc and the fluxes of Taurus, Ophiuchus, and ONC disks to the wavelength of 1.14~mm assuming a spectral index $\alpha=3$. 
Given the different methods used in the literature to derive the disk radius (see Appendix~\ref{Appendix:Radii}), 
we included disks whose radius has been measured with a power law\footnote{Note that \citet{Guilloteau2011} and \citet{Pietu2014} adopted a power-law radial dust temperature while we assumed a constant temperature.} or with a Gaussian fit, for which we used the relation\,C.4 of Appendix~\ref{Appendix:Radii} to obtain
the corresponding radii derived from a power-law fit.
The 1.14~mm scaled flux-size
relation shown in Fig~\ref{fmass}a reveals that for a given flux, disks in GGD\,27 cluster tend to be smaller than in Taurus-Auriga, with a clear deficit of disks with sizes $R_{\rm{disk}}\ga100$~au. Assuming the same conditions, \ie\ excluding disks of all samples below our sensitivity limit (dotted line in Fig~\ref{fmass}a), we obtained that the median radius in GGD\,27 is 34~au whereas that derived for Taurus and Ophiuchus disks 
is 92~au and 42~au, respectively. If we include the sample of Taurus-Auriga disks from \citet{Tripathi2017}, whose size has been measured with a different method (see Appendix~\ref{Appendix:Radii}), we still found that disks in the GGD\,27 cluster are smaller than in Taurus-Auriga. Recently, \citet{Maury2019} show that the youngest and more embedded Class~0 objects have relatively smaller disk ($R<60$~au for 75\% of their sample). This result is compatible with our hypothesis that disks in GGD\,27 are associated with embedded YSO.
We also compared our results with ONC \citep{Eisner2018} and obtained that the median radius of ONC disks is 44~au, similar to GGD\,27 and Ophiuchus. 
Therefore, we do not find significant differences between disks formed in cluster environments, such as GGD\,27 and ONC, and the Ophiuchus sample, which forms stars in isolation. Only disk population of Taurus-Auriga differs from the rest. This difference could be explained in terms of evolutionary effects or viscous evolution, since disks in Taurus are associated with Class~II YSOs, while the sample of disks in Ophiuchus and GGD\,27 contains objects in an earlier evolutionary stage.
However, we cannot rule out other processes such as different initial conditions or intrinsically different viscous timescales. Further observations toward a larger sample of disks in different cluster environments are needed to fully investigate the effects of a cluster on the disk population.

%--------------------------------------------------------
%-------------------------------------------------------

Given the clustered nature of GGD\,27 we might expect that its disk population could be affected mainly by two environmental factors: the high stellar density increases the probability of tidal/dynamical interactions, and second, the presence of a massive B star towards the cluster center may heat and photoevaporate disks of the surrounding low-mass cluster members. Observations and theoretical findings show that the mass-loss rate of disks due to external photoevaporation of the incident UV radiation field from the massive star depends on the distance of the disk to the central massive star \citep[\eg][]{Hollenbach1994,Ansdell2017}. Figure~\ref{fmass}b presents the disks masses as a function of the projected distance from the most massive object in the GGD\,27 cluster (\ie\ IRAS\,18162$-$2048, associated with MM1). Contrary to ONC, where there is a significant environmental impact with evidence of disk mass depletion at scales $<0.03$~pc \citep[][see also Fig.~\ref{fmass}b]{Mann2014,Eisner2018}, in GGD\,27 we found two massive disks (compared to ONC; ALMA\,7 and ALMA\,8) 
located close ($<$0.005~pc) to the massive star, similar to the case of the NGC\,2024 cluster \citep{Mann2015}. This could be explained either by the GGD\,27 cluster being younger than ONC, which is supported by the dynamical timescale of the radio jet ($\sim10^4$~years; \citealt{Masque12}) and the fact that the massive protostar has not yet developed a large HII region, or because the ionizing flux from GGD\,27 (B0-type star) is much weaker than in ONC (O6-type star).

In Fig.~\ref{fmass}c we show the disk's radius as a function of the projected distance from the massive GGD\,27 protostar. Interestingly, most of the disk population is located at distances $>0.02$~pc and only two disks are found in the immediate vicinity of the massive star, which could indicate that stars close to the massive star (\ie\ close to the cluster center) may not be capable to form disks. There is a clear lack of disks for distances between 0.006~pc and 0.02~pc, which is also observed in ONC but for smaller distances (see Fig.~\ref{fmass}b).
On the other hand, only for distances $>$0.04~pc disks appear 
larger and more massive than those located close to MM1 (Fig.~\ref{fmass}b and c). This result could indicate that dynamical interactions far from the cluster center are weaker, resulting that larger disks can easily form, even though the small disks ($R_{\mathrm{disk}}<50$~au) found at distances $>0.02$~pc could be the result of disk truncation due to interactions. 
However, a complete survey of the GGD\,27 disk sample (currently probably observed in its high-mass end with a mass sensitivity of 0.002~\msun) is needed to fully confirm both the absence of disk mass depletion and the deficit of disks close to the massive protostar since we cannot exclude the presence of faint and small disks.

In summary, the combination of sensitivity and spatial resolution leads us to the serendipitous discovery of a cluster of 25 continuum sources detected at 1.14~mm, most of them previously unknown. These 1.14~mm sources trace the dust emission of disks, with masses typically in the range 0.003-0.05\msun, surrounding Class~0 and Class~I low-mass protostars. 
Excluding the most massive, which is associated with MM1, the median value of the disk radius is 34~au, smaller than the median obtained for the Taurus disks but comparable to the results found in ONC and Ophiuchus, consistent with the fact that disks in Taurus are associated with more evolved YSOs.
The lack of disks close to the massive star GGD\,27\,MM1 suggests that in cluster environments, protostars close to the massive star may no be able to maintain their disks for a long time.
In the GGD\,27 cluster we only find large (and relatively massive) disks ($R_{\mathrm{disk}}>100$~au) at distances $>0.04$~pc from the massive star.
On the other hand, disks in the GGD\,27 cluster do not display evidence of external photoevaporation due to the massive star. However, observations with higher sensitivity are needed to fully support these results. 
This study shows the potential of ALMA to unveil a disk population at millimeter wavelengths, whose emission appears completely hidden in the near-infrared due to the strong emission arising from the massive star. This will open a new window for ALMA to investigate the properties of protostellar disks in deeply embedded young cluster environments located at large distances.

\begin{acknowledgements}
We sincerely thank the referee for his/her valuable comments that helped to improve this paper. 
This paper makes use of the following ALMA data: ADS/JAO.ALMA\#2015.1.00480.S. ALMA is a partnership of ESO (representing its member states), NSF (USA) and NINS (Japan), together with NRC (Canada) and NSC and ASIAA (Taiwan) and KASI (Republic of Korea), in cooperation with the Republic of Chile. The Joint ALMA Observatory is operated by ESO, AUI/NRAO and NAOJ.
G.B., J.M.G., R.E., G.A., N.AL, M.O., and J.M.T. are supported by the MINECO (Spain) AYA2014-57369-C3 and AYA2017-84390-C2 grants (co-funded by FEDER). S.C. acknowledges support from DGAPA, UNAM and CONACyT, M\'exico.  
\end{acknowledgements}

\bibliographystyle{aa}
\bibliography{girart}

\begin{appendix}
\section{Table}\label{Appendix:Table}
Table~\ref{tsources} lists the millimeter continuum sources detected in the GGD\,27 region as explained in Sect.~2. We report the source coordinates, the local rms measured in a region locally around each millimeter source (as explained in Section~\ref{Sec:Obs}), the peak intensity, the flux density obtained by fitting a two-dimensional Gaussian function, and the peak brightness temperature. We also report the counterpart identification at other wavelengths based on the work carried out by \citet{Pravdo04,Pravdo09}, who cross-correlated the X-ray sources with sources from the 2MASS and USNO B1 catalogs and with the \textit{Spitzer}/IRAC detections, the radio continuum studies from the literature, and the high angular resolution ($\sim0\farcs3$) JVLA observations at 1~cm (project ID: VLA15A-132). Finally, we list the infrared spectral classification, when available, based on the work of \citet{Pravdo09}.\\

\begin{landscape}
\begin{table}
\caption{ALMA 1.14~mm continuum sources towards GGD\,27.}
\begin{footnotesize}
\begin{tabular}{llccccccccccc}
\\
\hline
&  &
\multicolumn{1}{c}{$\alpha$(J2000)$^a$} &
\multicolumn{1}{c}{$\delta$(J2000)$^a$} &
$\sigma^b$ & $I_{\nu}$ & $S_{\nu}$ &T$_{\rm{b}}$ &
\\
\cline{3-4}
&  
&
\multicolumn{1}{c}{$18^{\rm h} 19^{\rm m}$} &
\multicolumn{1}{c}{$-20$\grau$47'$} &
($\mu$Jy & (mJy &(mJy) &(K) &
Counterpart & 
Infrared &
\\
id. &
IAU nomenclature& 
($^{\rm s}$) &
($''$) & 
beam$^{-1}$) & beam$^{-1}$) & &
& at other wavelengths &Spec. Class.$^c$ & References 
\\
\hline 
ALMA 1  & J181911.48-204722.8 & 11.4834 & 22.821	& 37 & \phn0.47 & \phnn0.47\Mm0.07 &\phnn5.45   &X-ray source: 6a &II &1 \\
&&&&&&&&2MASS J18191146-2047229 &\\
ALMA 2  & J181911.80-204740.5 & 11.8025 & 40.522	& 33 & \phn0.95 & \phnn1.50\Mm0.09 &\phn11.04  &CXOPTM J181911.8–204740 &III &1,2\\
&&&&&&&&(X-ray source: 13) & \\
&&&&&&&&2MASS: 18191184–2047406 \\
ALMA 3  & J181911.84-204743.6 & 11.8431 & 43.638	& 40 & \phn1.77 & \phnn4.57\Mm0.18 &\phn20.57	& & \\
ALMA 4  & J181912.09-204727.4 & 12.0868 & 27.356	& 24 & \phn0.61 & \phnn0.66\Mm0.05 &\phnn7.09 & \\
MM1/ ALMA 5$^d$ & J181912.09-204730.9 & 12.0950 & 30.952& 50 & \phn46.00 & 351.30\Mm0.33	&534.65 &X-ray source: 9a &I &1, 3, 4  \\
&&&&&&&&Radio source \\
&&&&&&&&IRAS\,18162$-$2048 \\
ALMA 6  & J181912.11-204725.6 & 12.1141 & 25.579	& 24 & \phn0.25 & \phnn0.28\Mm0.05 &\phnn2.91	& \\
ALMA 7  & J181912.12-204730.3 & 12.1212 & 30.322	& 62 & \phn0.69 & \phnn1.06\Mm0.23 &\phnn8.02	& \\
ALMA 8  & J181912.14-204730.7 & 12.1415 & 30.734 	& 50 & \phn1.03 & \phnn1.44\Mm0.19 &\phn11.97 	& \\ 
ALMA 9  & J181912.17-204728.2 & 12.1705 & 28.154	& 29 & \phn0.50 & \phnn0.56\Mm0.05 &\phnn5.81 &H$_2$O maser & &9 \\
ALMA 10  & J181912.22-204742.0 & 12.2216 & 41.992	& 33 & \phn0.34 & \phnn0.38\Mm0.06 &\phnn3.95 & 	\\
ALMA 11 & J181912.22-204728.7 & 12.2236 & 28.703	& 25 & \phn0.21 & \phnn0.32\Mm0.07 &\phnn2.44 &2MASS: J18191220-2047297   \\	
&&&&&&&&radio source: VLA\,2$^{e}$ &&5 \\
ALMA 12 & J181912.40-204722.8 & 12.4022 & 22.827	& 30 & \phn0.24 & \phnn0.24\Mm0.05 &\phnn2.79 &Radio Source CS3 &&6  \\
MM2(W)/ ALMA 13 & J181912.44-204727.6 & 12.4431 & 27.611	& 29 & \phn2.46 & \phnn2.82\Mm0.06 &\phn28.59	&Radio source & &4   \\   

ALMA 14 & J181912.46-204734.2 & 12.4587 & 34.175	& 23 & \phn0.93 & \phnn1.11\Mm0.05 	&\phn10.81 & 	\\
ALMA 15 & J181912.46-204735.1 & 12.4626 & 35.075 	& 23 & \phn0.97 & \phnn1.00\Mm0.13 	&\phn11.27 & 	\\
ALMA 16 & J181912.48-204735.3 & 12.4757 & 35.332	& 23 & \phn1.07 & \phnn2.70\Mm0.17	&\phn12.44 &CXOPTM J181912.4-204733 &I &1, 2\\
&&&&&&&&X-ray source: 10 & \\
MM2(E)/ ALMA 17  & J181912.51-204727.4 & 12.5105 & 27.438 	& 39 &15.15& \phn33.55\Mm0.43 &176.09	&Radio source: VLA\,3 & &5, 4, 7, 8 \\
&&&&&&&&H$_2$O maser & \\
ALMA 18 & J181912.58-204721.5 & 12.5764 & 21.481	& 35 & \phn0.42 & \phnn0.46\Mm0.07	&\phnn4.88 &  \\
ALMA 19 & J181912.59-204737.8 & 12.5947 & 37.813	& 31 & \phn0.30	& \phnn0.37\Mm0.06	&\phnn3.49 &  \\
ALMA 20 & J181912.85-204724.1 & 12.8533 & 24.051	& 38 & \phn0.58 & \phnn1.86\Mm0.17	&\phnn6.74 &  \\  
ALMA 21 & J181912.92-204739.2 & 12.9212 & 39.345	& 51 & \phn0.74 & \phnn1.21\Mm0.13 &\phnn8.60 &X-ray source: 11 &&1  \\ 
ALMA 22 & J181912.95-204735.5 & 12.9480 & 35.548	& 40 & \phn0.36 & \phnn0.40\Mm0.09 &\phnn4.18 &  \\
\hline
\multicolumn{3}{l}{Tentative detections: } \\
ALMA 23 & J181911.10-204721.3 & 11.1016 & 21.301 & 89 & \phn0.54 & \phnn1.22\Mm0.21 &\phnn6.28 & \\
ALMA 24 & J181912.80-204740.6 & 12.8040 & 40.637 & 57 & \phn0.29 & \phn0.44\Mm0.07 &\phnn3.37 &\\
ALMA 25 & J181913.14-204739.2 & 13.1354 & 39.148 & 77 & \phn0.47 & \phn0.44\Mm0.10 &\phnn5.46 & \\
\hline 
\end{tabular}
\\
{\em $^a$}  The position uncertainties are 3~mas or better. \\
{\em $^b$}  Local rms measured in a region close to the millimeter source. \\
{\em $^c$} Infrared Spectral Classification from \citet{Pravdo09}. \\
{\em $^d$}  Peak intensity and flux density obtained after subtracting the compact source component (assumed to be free-free emission) detected in \citet{Girart2018}. Therefore, these values correspond to the contribution of the dust emission.\\
{\em $^e$} The radio continuum source VLA\,2 is $1\farcs4$ offset from ALMA\,11.  
\\
\textbf{References:} (1): \citet{Pravdo09}; (2): \citet{Pravdo04}; (3): \citet{Kurtz1994};(4) JVLA data from project VLA15A-132; (5): \citet{Gomez95}; (6): \citet{Yamashita1991}; (7): \citet{Gomez03}; (8): \citet{Marti1999}; (9) \citet{Kurtz2005} \\
\\
\label{tsources}
\end{footnotesize}
\end{table}
\end{landscape}

\section{Background sources}\label{Appendix:Backrground}

The analysis carried out in this paper assumes that the detected millimeter sources are tracing disks of dust around YSOs. This assumption is based on the small size of the sources: a few 100~au or less, if they belong to the GGD\,27 region. Alternatively, given the excellent sensitivity of ALMA, it could be feasible that some of them could be tracing dust emission from background extragalactic sources. To estimate how many of these could appear within the field of view, we used as a reference the deep 1.1~mm ALMA observations carried out by \citet{Hatsukade16} over a 2~arcmin$^2$ field. According to that work, for a flux density above than 0.2~mJy (\ie\ above the $8\sigma$ threshold for a source detection), the number of sources per square degree is $38^{+24}_{-9}$\,$\times10^3$.
Figure~\ref{fig:FOV} shows a field of view of $\sim0.16$~arcmin$^2$. That means that the number of extragalactic sources above the aforementioned flux level would be 2$\pm1$.
This represents an upper limit given that we considered the lowest rms of our ALMA image (see Table~\ref{tsources}). Therefore, we expect no more than one extragalactic sources in our ALMA field of view.

\section{Disk masses and radii}\label{Appendix:Radii}

We estimated the mass of gas and dust of the disk population of the GGD\,27 protostellar cluster assuming that the 1.14~mm dust continuum emission is optically thin and the temperature distribution is uniform. We adopted a gas-to-dust ratio of 100 and the opacity law of \citet{Beckwith1990} with a dust opacity coefficient of $\kappa_\nu=2.63$\cmd\,g$^{-1}$ at 263~GHz and a spectral index $\beta=1$. We assigned a dust temperature of 20~K to most of the millimeter sources in our sample based on the median temperature for the Taurus disks \citep{Andrews2005}. We made the caveat that using $T_{\rm{d}}=10$~K and $T_{\rm{d}}=30$~K instead, the masses would be a factor of about $8/3$ and $2/3$ those obtained using $T_{\rm{d}}=20$~K, respectively. The uncertainty in the masses due to an uncertainty in the dust opacity is a factor of two. For the two most massive sources, we adopted the more appropriate values of $T_{\rm{d}}=109$~K for MM1 and $T_{\rm{d}}=35$~K for MM2(E) and MM2(W) as found in \citet{FernandezC11} and \cite{Qiu09}. The derived masses for MM1, and specially for MM2(W) and MM2(E) are smaller than the one derived from the SMA with a larger beam \citep{FernandezL11_MAM}. This is probably due to presence of extended emission from the core that is being filtered out by ALMA. It is important to mention that the derived masses are most likely a lower limit given the assumption of optically thin emission, which may not hold for all disks in GGD\,27. For instance, \citet{Girart2018} found that the disk associated with ALMA\,5/MM1 is optically thick.

In order to determine the characteristics of the disks, we modeled the emission of the disks seen at an inclination angle $i$ ($i=0$\grau\ for a face-on disk and $i=90$\grau\ for an edge-on disk) with an intensity that is a power-law of the radius ($I_\nu\propto r^{-q}$), up to a maximum radius $R_\mathrm{disk}$, convolved with a Gaussian beam with a HPBW equal to that of the observation. The model depends on a maximum of 8 parameters:
the position of the disk center, $(x_0, y_0)$;
the position angle of the disk axis, $PA$;
the disk inclination, $i$;
the disk radius, $R_\mathrm{disk}$;
the intensity power-law index, $q$;
the peak intensity, $I_\nu^\mathrm{peak}$;
and
the background intensity, $I_\nu^\mathrm{bg}$.

The fitting procedure was similar to that used in other fittings by \citet{Palau14} and \citet{Estalella17}. The fit was performed by sampling the space parameter of dimension up to 8 using a Halton pseudo-random sequence, to find the minimum value of the rms fit residual. The details of the method and the derivation of the parameter uncertainties can be found in \citet{Estalella17}.

The value of the power-law index resulted ill-determined for disks with a low signal-to noise ratio (SNR) or poorly angularly resolved. Thus, we decided to use the same value of $q$ for all the disks, $q= 1.1$, except for ALMA\,17. This value was derived from the fit to the disk with the highest SNR and well resolved by the observation, ALMA\,3 (see Fig.~\ref{fig:Sources}). For ALMA\,17 we fitted the radial intensity profile with the power law index $q$ as a free parameter, and obtained $q=1.46$. In addition, we used a constant background intensity $I_\nu^\mathrm{bg}= 0$ for all the disks, except for ALMA\,7 and ALMA\,8, where a cutoff of 0.20~\mJy\ and 0.15~\mJy\ was used, respectively.
When the fit gave a value of the inclination consistent with a face-on disk, a second fit was performed forcing the inclination to be zero, and the position angle to be an arbitrary constant value. Thus, in general we fitted 6 parameters for inclined disks, and 4 for face-on disks. Finally, when the error in the fitted radius was equal or larger than the value itself, we considered that the disk was unresolved, and only an upper limit to $R_\mathrm{disk}$ was given. Table~\ref{tdiskpar} shows the disk parameters inferred by our model. It is important to point out that the disk radii have been obtained using the ALMA image that excludes baselines shorter than 300~$k\lambda$. However, the derived disk sizes do not change significantly if all visibilities are included, being always within the uncertainty of the 2D Gaussian fit.

%--------------------------------------------------------
\begin{figure}[!t]
\begin{center}
\begin{tabular}[t]{cc} 
\epsfig{file=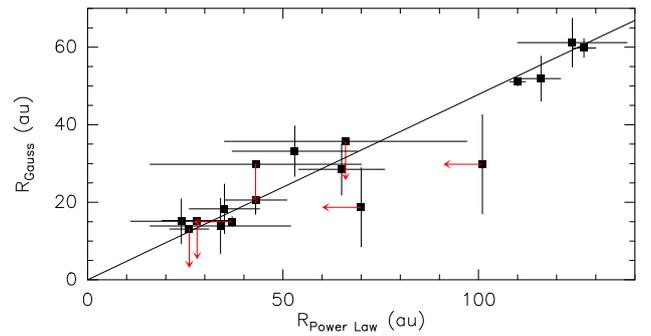,scale=0.45,angle=0}
\end{tabular}
\caption{Comparison between the disk radii obtained with a Gaussian fit and adopting a power law with an index of $q=1.1$ for the sample of disks (Table~\ref{tdiskpar}). Red arrows indicate an upper limits. The solid lines shows the linear regression from the data, which indicates a relationship of $R_{\rm Gauss}=0.48\pm0.04 \times R_{\rm Power\,law}$.
}
\label{fradii}
\end{center}
\end{figure}
%-------------------------------------------------------

It is interesting to compare the radii derived from this model and the radii 
derived from a Gaussian fit to the intensity observed. A Gaussian fit can be 
understood as a fit that reproduces the moments up to order two of the 
observed intensity. Let us call $R_G$ the radius, i.e. the half width at half maximum, 
 derived from the Gaussian fit,
\begin{equation}
G(r) = \frac{\ln2}{\pi} \frac{F_\nu}{R_G^2} \exp\left(-\ln2\frac{r^2}{R_G^2}\right),
\end{equation}
where we assume that the disk is centered at $r=0$ (first order moment),
$F_\nu$ is the flux density (zeroth order moment), and 
the normalized second order moment in any direction of the Gaussian is 
\begin{equation}
\mu_2= \frac{\int\!\!\int_{-\infty}^{+\infty} x^2 G(r)\,dx\,dy}{\int\!\!\int_{-\infty}^{+\infty} G(r)\,dx\,dy}= \frac{R_G^2}{2\ln2}.
\end{equation}
The model used in this paper assumes an intensity that is a power law of the 
radius, $I=I_0 r^{-q}$ up to a radius $R_\mathrm{disk}$. The normalized second-order moment 
of the power-law intensity can be easily computed and results in
\begin{equation}
\mu_2= \frac{1-q/2}{4-q} R_\mathrm{disk}^2.
\end{equation}
If we compare this result with the Gaussian fit, we find, for the value of $q$ 
found in this work, $q=1.1$,
\begin{equation}
R_G= \left[2\ln2\frac{1-q/2}{4-q}\right]^{1/2}R_\mathrm{disk}= 0.46\,R_\mathrm{disk}.
\end{equation}
Thus, radii derived from a Gaussian fit are $\sim 2.2$ times smaller than the radii derived from a power-law fit with a power-law index $q=1.1$.  Indeed, from our data the comparison between the measured radius adopting the aforementioned power law with respect to the radius derived from a Gaussian fit shows an excellent agreement with this result (see Fig.~\ref{fradii}).  In three of the disks, ALMA\,7, ALMA\,11 and ALMA\,13, the power law fit gives only upper limits. In these case, we used the results of a Gaussian fit by applying the above correction factor.

The above two methods used to define disks radius are not the only ones found in the literature. For cases where the emission is well resolved, a surface density that follows a modified power with an exponential decay at larger radius: $\Sigma (r) \propto (r/R_0)^{-\gamma} \, \mathrm{exp}(-((r/R_{\rm c})^{2-\gamma})$ is also used \citep[e.g.,][]{Andrews2009}. In these cases the adopted disk radius is $R_{\rm c}$, the ``transition'' radius between the power law and the exponential zones. On the other hand, in the compilation done by \citet{Tripathi2017}, they use a different definition of radius, the so-called effective radius, $R_{\rm eff}$, that corresponds to the radius where 68\% of the total flux is located within this radius. We have compiled a list of sources where the disk radius has been derived from a simple power law fit and also from either (or both) $R_{\rm c}$ and $R_{\rm eff}$ \citep{Guilloteau2011,Isella2010,Pietu2014,Tripathi2017}. We found that for disks with a power law index around 1, the radius obtained with a simple power law is in average about two times larger than the $R_{\rm c}$ and $R_{\rm eff}$, which is a similar result as the one found with respect a Gaussian fit.

\section{Statistical analysis of the cluster}\label{appendix:cluster}

Although features such as clustering or gradients in the density of sources can be noticed by eye, some statistical tools have been developed to study the spatial distribution of the members of a given cluster. 
In this section we first  determine (\ref{appendix:distribution} and \ref{appendix:mst}) whether the GGD\,27 cluster of protostellar disks is centrally distributed or is clustered. To achieve this we follow the formalism of \citet[][hereafter CW2004]{2004Cartwright} by using the minimum spanning tree (MST) and the distribution function of all the possible separations between the cluster members\footnote{We consider all the sources reported in Table~\ref{tsources} as cluster members, including the tentative detections ALMA\,23 to ALMA\,25.}. 
Second, we analyze (\ref{appendix:segregation}) the mass segregation \cite[see][]{2009Allison}. This analysis is based on the mass measured from dust continuum ALMA observations, which mainly arises from disks surrounding YSOs. We would like to point out that the infrared \textit{Spitzer} and X-ray \textit{Chandra} data show that there is a larger stellar population associated with the GGD\,27 cluster \citep{Qiu08,Pravdo04,Pravdo09}. However, we focused the following analysis toward the center of the cluster core (i.e., in a $\sim$0.15~pc region) close to the massive protostar GGD\,27~MM1.

%--------------------------------------------------------
\begin{figure*}[!t]
\begin{center}
\begin{tabular}[t]{cc} 
\epsfig{file=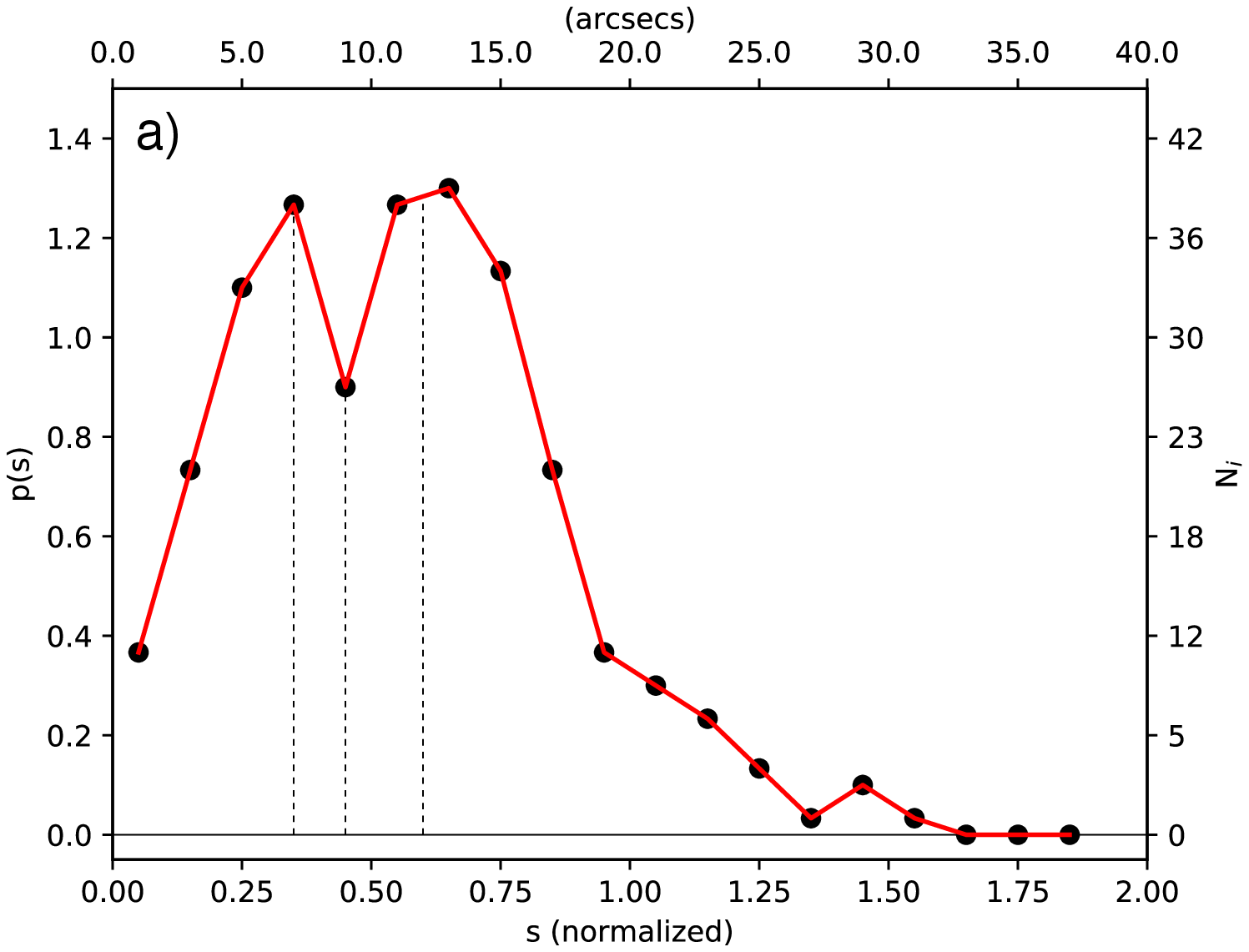,scale=0.66,angle=0} &
\epsfig{file=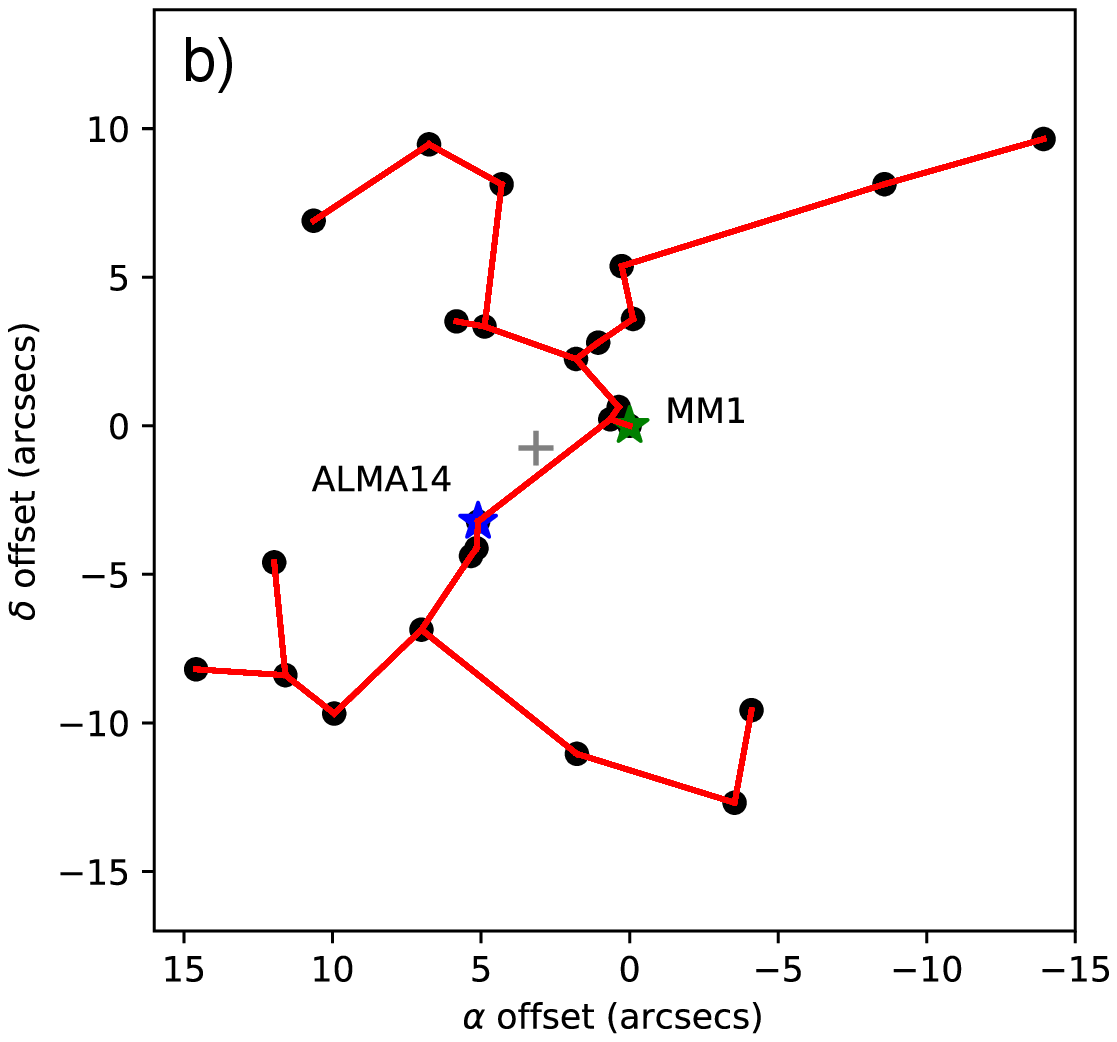,scale=0.66,angle=0} 
	\end{tabular}
\caption{\textit{a)} Distribution function of separations between protostellar disks in the GGD\,27 cluster. The quantity $p(s)~\Delta s$ is the probability that the projected separation between two randomly selected disks of the cluster is within $s$ and $s+\Delta s$. Separations $s$ are normalized to the cluster radius ($20''$) in the bottom axis, while the top axis extends up to twice this radius. The radius is determined as the distance from the average position of all the cluster members to its furthest protostellar disk. Dotted vertical lines mark the two peaks and the dip in the $p(s)$ distribution. The right axis represents the number of pairs of stars within each separation bin $\Delta s$. \textit{b)} Minimum spanning tree for the ALMA detections of the GGD\,27 cluster. The positions of the cluster members are given with respect to the position of MM1 (ALMA\,5) of Table~\ref{tsources}. MM1 and ALMA\,14 are marked with colored stars. The grey cross marks the average position of the GGD\,27 sources.}
\label{fmst}
\end{center}
\end{figure*}
%-------------------------------------------------------

\subsection{Distribution function}\label{appendix:distribution}

Following CW2004 we derived the distribution function, $p(s)$, of the projected separations ($s$) between pairs of cluster stars. By counting the number of pairs $N_i$, we obtain the probability that the separation between whatever two stars of the cluster is in the interval $(s, s + \Delta s)$. We used the normalized expression:
\begin{equation}
p(s)\Delta s=\frac{2 N_i}{N_\mathrm{tot}(N_\mathrm{tot}-1)}\quad 
\end{equation}
where $N_\mathrm{tot}$ is the total number of stars in the cluster (25 in our ALMA observations of GGD\,27). $p(s)$ is the number of separations inside a range of width $\Delta s$, divided by the total number of separations in the cluster ($25\cdot24/2=300$). We made 20 bins using $\Delta s=2 R_{\rm{GGD27}} / 20=2\arcsec$ which provides an adequate sampling of the dataset. Note that the angular resolution is about 40~mas, and we defined the cluster radius $R_{\rm{GGD27}}=20\arcsec$ as the maximum separation between a star in the cluster and the average position of all the cluster stars.

Figure~\ref{fmst}a shows the distribution function $p(s)$ of the separations measured for the protostellar disks of the GGD\,27 cluster as a function of the separation $s$. CW2004 demonstrated that for clusters showing a smooth volume density profile, $n\propto r^{-\alpha}$, the distribution function has only one maximum, while that of GGD\,27 presents two peaks separated by $6''$ ($\sim0.04$~pc), indicative of subclustering and possible fractal structure. 
The first peak is located at a distance of $6''$ or 0.04~pc, which approximately corresponds to the distance of three main groups in GGD\,27: MM1 and ALMA\,7-8 with ALMA\,14-16, MM1 and ALMA\,7-8 with MM2(W) and MM2(E), and MM2(W) and MM2(E) with ALMA\,14-16.  The dip between the two peaks corresponds to a separation of about $8.5\arcsec$ ($\sim0.06$~pc) and it is significant as long as the binning in the $p(s)$ versus $s$ plot is not larger than $2\arcsec$ ($\sim0.01$~pc), which is approximately one third of the separation between the two peaks in that plot. The second peak in the plot corresponds to separations about $12\arcsec$ ($\sim0.08$~pc). We note that the decline for $s>1$ is due to the ALMA primary beam. The sampling of separations is probably good enough up to this distance within our field of view, but note that these results can change if fainter sources are detected (through deeper observations) or a more spread out mosaic is carried out and new sources are detected outside of the current ALMA primary beam.
Finally we estimate the mean separation between the GGD\,27 disks ($11\farcs8$, about 0.08~pc) and normalize it to the cluster radius, $\bar{s}=0.59$. We will use $\bar{s}$ in the next section to derive the diagnostic parameter $Q$.

\subsection{$Q$ parameter}\label{appendix:mst}

The $Q$ parameter was introduced by CW2004 to discern between clusters with a gradual decrement on the stellar density (centrally peaked clusters) and those showing a self-similar or subclustering appearance \citep[fractal, see also][]{Parker2018}. In order to use this formalism the minimum spanning tree (MST) of the cluster has to be constructed. The MST measures the shortest path length connecting all points in a given sample, without including closed loops \citep{1956Kruskal}. Figure~\ref{fmst}b presents the MST for the GGD\,27 cluster. It shows two overdensities of sources around MM1 and ALMA\,14 locations (colored stars), separated by $\sim6\farcs5$ (0.05~pc), roughly matching the separations of the first peak in Fig.~\ref{fmst}a.

Following CW2004 and once the MST of the GGD\,27 cluster has been built, we derived the normalized mean edge length, $\bar{m}=$0.40, defined as the sum of the MST paths linking all the stars divided by the normalization factor $\sqrt{N_{\mathrm{tot}}\,A}/(N_{\mathrm{tot}}-1)$, 
where $A$ is the cluster projected area.
Having $\bar{m}$ and $\bar{s}$, we now obtain the value for the parameter $\bar{Q}=\bar{m}/\bar{s}=$0.68. This parameter serves as a diagnostic to distinguish between a cluster with a smooth overall radial density gradient of sources ($\bar{Q}>0.8$) and one with a multiscale fractal subclustering ($\bar{Q}<0.8$). GGD\,27 results to be of the second type, and 
moreover, given its $\bar{Q}$ value we estimate that it has a notional fractal dimension $D\simeq$2.2 (see Fig.~5 in CW2004), which stands for a moderate subclustering ($D$ ranges between 1.5, for strong subclustering, and 3, for no subclustering). We note that by using this method it cannot be decided if a cluster is truly fractal in the sense that it replicates structures at different spatial scales (\ie\ if it is hierarchically self-similar). Low values of $\bar{Q}$, such as the one found in GGD\,27, are expected in young clusters, before any primordial subclustering washes out as the cluster acquires a radial density profile due to dynamical relaxation. Therefore, this supports the youthfulness of the GGD\,27 cluster.

\subsection{Mass segregation}\label{appendix:segregation}

For our second statistical analysis, let us define the Mass Segregation Ratio \citep[$\Lambda_\mathrm{MSR}$, ][]{2009Allison}:
\begin{equation}
\Lambda_\mathrm{MSR}=\frac{L_\mathrm{norm}}{L_\mathrm{massive}}\pm\frac{\sigma_\mathrm{norm}}{L_\mathrm{massive}}
\end{equation}
where $L_\mathrm{massive}$ is the summation of the path-lengths in the MST of the $N$ most massive sources.  $L_\mathrm{norm}$ is the average path-length of the MST of $N$ sources picked randomly in the cluster.
We take an average over 1000 random sets of $N$ sources in the cluster to estimate $L_\mathrm{norm}$ and derive its statistical deviation $\sigma_\mathrm{norm}$ along the 1000 runs \citep{2017Gavagnin}. If $\Lambda_\mathrm{MSR}$ is greater than unity, the path-length of the $N$ most massive sources is shorter than the average path-length of $N$ randomly chosen sources in the cluster, indicating a concentration of massive sources with respect to the random sample and hence a mass segregation.

To build up the plot in Fig.~\ref{fcluster} we estimated $\Lambda_\mathrm{MSR}$ for $N=2, \ldots,25$. We have removed ALMA\,1 and ALMA\,2 from this analysis since they are associated with a Class~II and a Class~III YSO, respectively \citep[][see also Table~\ref{tsources}]{Pravdo09}. In addition, \citet{Pravdo09} classified MM1 and ALMA\,16 as Class~I protostars. The remaining 21 ALMA dust continuum sources do not have an infrared counterpart and are not detected in X-rays \citep[see][]{Pravdo09}, suggesting that these disks are most likely associated with Class~0/I protostars. In addition, the very high extinction that may hidden X-ray sources only affects the north-eastern region \citep[see the NH$_3$ emission map of][]{Gomez03}.
Thus, our analysis accounts only for the youngest objects of the GGD\,27 cluster. Figure~\ref{fcluster} shows a dominant $\Lambda_{MSR}$ value of about one, which means no mass segregation or random distribution regarding the protostellar disk masses.
Only the two most massive protostellar disks, MM1 and MM2(E), which are located at the center of the gravitational potential of the cluster \citep[e.g., see][]{Gomez03}, appear segregated from the rest, with a separation about 1.7 times shorter than the average distance between two sources picked up at random from the cluster. This result, together with the low value of the $\bar{Q}$ parameter, is in agreement with the expected youth of the cluster (older clusters are dynamically relaxed, implying that mass segregation has taken place already). 
Thus, it is probable that the disks detected are associated with protostars (Class~0 and Class~I YSOs), since disks in Class~II and Class~III YSOs would be too faint to be detected with our sensitivity. For instance, \citet{Plunkett2018} find that none of the Class~II sources and most of the Flat spectrum YSOs identified with \textit{Spitzer} by \citet{Dunham2015} in the Serpens South protocluster (located at a distance of 436~pc) is detected at millimeter wavelengths with ALMA with a sensitivity of 1.5~\mJy. At the distance of GGD\,27 this sensitivity would correspond to a flux density of 145~$\mu$Jy, hence a Class~II YSOs in GGD\,27 would be well below our 3$\sigma$ rms noise level. 
On the other hand, prestellar cores are too extended so they are completely filtered out by the interferometric observations. Thus, the age problem is minimal in our sample of millimeter sources, which may include disks from Class~0 and Class~I protostars, and only two disks associated with a Class~II (ALMA\,1) and a Class~III (ALMA\,2) YSOs, which have been removed from the mass segregation analysis.
However, we should make some caveats regarding the derived masses. The resulting mass estimates for the protostellar disks may not correlate with the masses of the protostars. While there is an observational trend between the disk mass and the stellar mass \citep[see \eg][]{Pascucci2016} the large scatter found prevented us to adopt this relation to derive the stellar mass. Nevertheless, the correlation between protostellar masses and protostellar disk masses can be roughly true for young Class~0 and Class~I protostars, for which the derived mass contains most the mass of the whole system \citep{2000Andre}. Therefore we are confident that our analysis indicates that the spatial distribution of the protostellar disk population do not depend on the protostellar disk masses, and hence the derived masses reflect the lack of primordial mass segregation.

\end{appendix}

\end{document}